\documentclass[a4paper,fleqn,usenatbib]{mnras}
\usepackage{graphicx}

\usepackage[utf8]{inputenc}
\usepackage[T1]{fontenc}
\usepackage{amsmath,amssymb}
\usepackage{color}
\usepackage{newtxtext,newtxmath}

\DeclareMathOperator{\sech}{sech}

\title{Spectral signatures of recursive magnetic field reconnection}

\author[A. Tenerani and M. Velli]
{A. Tenerani,$^{1}$ \thanks{Email address for correspondence: Anna.Tenerani@austin.utexas.edu}, M. Velli,$^2$ \\
$^1$Department of Physics, The University of Texas at Austin,~TX\\
$^2$Department of Earth, Planetary, and Space Sciences, University of California, Los Angeles,~CA
}

\begin{document}

\maketitle

\begin{abstract}
We use 2.5D Magnetohydrodynamic  simulations to investigate the spectral signatures of the nonlinear disruption of a tearing unstable current sheet via the generation of multiple secondary current sheets and magnetic islands.  During the nonlinear phase of tearing mode evolution, there develops a regime in which the magnetic energy density shows a  spectrum with a power-law close to $B(k)^2\sim k^{-0.8}$. Such an energy spectrum is found in correspondence of the neutral line, within the diffusion region of the primary current sheet, where energy is conveyed towards  smaller scales via a  ``recursive''  process of fast tearing-type instabilities. Far from the neutral line we find that magnetic  energy spectra evolve towards slopes compatible with the ``standard''  Kolmogorov spectrum. Starting from a self-similar description of the nonlinear stage  at  the neutral line, we provide a model that predicts a reconnecting magnetic field energy spectrum scaling as  $k^{-4/5}$, in good agreement  with  numerical results. An extension of the predicted power-law to generic current sheet profiles is also given and possible implications for turbulence phenomenology are discussed.  These results provide a step forward to understand the  ``recursive''  generation of magnetic islands (plasmoids), which has been proposed as a possible explanation for the energy release during flares, but which, more in general, can have an impact on the subsequent turbulent evolution of unstable sheets that naturally form in the high-Lundquist number  and collisionless plasmas found in most of the astrophysical environments.
\end{abstract}


\section{Introduction}

Understanding catastrophic processes of energy storage and  release via magnetic field reconnection requires, first and foremost, an understanding of under which conditions current sheets become unstable and then evolve nonlinearly.  In collisionless or weakly collisional plasmas like those found in many astrophysical environments, thin current sheets spontaneously arise as the result of the plasma  dynamics. For example, current sheets are thought to necessarily form in the solar corona as a result of the photospheric displacement of closed coronal loop field lines~\citep{parker_1972, rappazzo_2013}. Locally, the dissipation or instability of such current sheets is thought to produce heating and particle acceleration, while at larger scales, where topological invariants (helicity conservation) may play a role, current sheet evolution may be driven self-consistently in the eruption process leading to flares~\citep{flares}. Similarly, in the magnetosphere, a current sheet forms in the tail as a consequence of the external solar wind driving~\citep{otto_2015}. Numerical simulations of these processes have only began to reach the Lundquist and Reynolds numbers sufficient to study the formation of thin sheets and the triggering of their instabilities.

It is now well established that the behavior of current sheets changes dramatically with increasing macroscopic Lundquist number $S=Lv_a/\eta$, where $L$ is the (half) length of the sheet, $v_a$ the upstream Alfv\'en speed and $\eta$ the magnetic diffusivity. For low values of the Lundquist number, $S<S_c$, $S_c$ being a (non-universal) critical threshold around~$S_c\simeq10^4$~\citep{ni_2010, shi_2018}, stable Sweet-Parker-type current sheets can form where laminar, slow reconnection is sustained by plasma flows into and outwards from the sheet~\citep{sweet, parker_57}. For Lundquist numbers larger than $S_c$, however, Sweet-Parker sheets, whose  aspect ratio scales as $L/a\sim S^{1/2}$, are so thin that they become unstable to extremely fast  tearing~\citep{bisk}, with normalized growth rate  $\gamma\sim S^{1/4}$~\citep{tajima,loureiro_2007, bhattacharjee_2009}.  For example, an active region in the solar corona has a typical spatial scale $L$ of about $L\simeq 10^9$~cm, a magnetic field on the order of $B\simeq 50$~G, density $\rho\simeq 10^9$~cm$^{-3}$ and a temperature $T\simeq 10^6$~K  that give a macroscopic Lundquist number $S\simeq10^{13}$. At such large values of $S$, a Sweet-Parker sheet would be unstable on a  too short of a  timescale (less than a second). The opposite  happens instead for a current sheet which is macroscopic, which would be unstable on infinitely long timescale. 

On the other hand, a fast tearing instability can grow on an ideal, i.e., $S$-independent, timescale  ($\gamma\sim \mathcal{O}(1)$)  within current sheets whose aspect ratio  $L/a\sim S^{1/3}$---much smaller than the Sweet-Parker one~\citep{pucci}.  The fundamental implication of the existence of such an unstable mode, dubbed ``ideal'' tearing, is that the predicted  critical  aspect ratio provides an upper limit for current sheets that can naturally form before they disrupt due to the onset of a fast reconnecting mode.  In other words, a current sheet, during its  dynamical formation,  remains quasi-stable as long as its thickness is larger than critical (in which case the growth rate is infinitely small if $S\gg 1$), but it becomes unstable with a finite growth rate while approaching the critical thickness from above. Such a scenario  of the disruption of a forming current sheet, as well as the scaling with $S$ of the critical aspect ratio, has been  confirmed by Magnetohydrodynamic (MHD) numerical simulations of forming current sheets~\citep{anna, huang_2017}.   Although ``ideal'' tearing  was first discussed in the framework of resistive MHD, it was later  shown that the specific scaling of $L/a$ with plasma parameters may change depending whether viscous or kinetic effects are included, and also depending on the current sheet profile chosen~\citep{anna0, delsarto_2016, pucci_2017, pucci_2018}.   

The idea of a threshold aspect ratio at which magnetic reconnection is triggered on a timescale compatible with the ideal dynamics---i.e., on a timescale generally independent of the non-ideal terms---has been adopted in recent work aimed at incorporating the effect of the tearing instability, within anisotropic eddies, on the turbulent energy cascade~\citep{loureiro_2017, mallet}. Here we discuss a related problem, namely that of how magnetic reconnection mediates the formation of small scales,  by considering the nonlinear evolution of a (two-dimensional) single current sheet.  Instead of a Sweet-Parker current sheet, which has been often considered as the archetype of current sheets, we consider a current sheet at the critical thickness defined above, therefore unstable to the ``ideal'' tearing mode.   We investigate via  MHD simulations how the nonlinear, recursive onset of tearing-type instabilities within ever smaller current sheets provides a channel to convey energy across scales all the way down to the dissipative ones. 

Numerical studies of magnetic reconnection in large aspect ratio sheets, implementing different initial conditions and adopting different plasma descriptions, have shown that  X-points collapse and evolve into  secondary current sheets during the nonlinear stage of the instability~(e.g., \cite{malara, jemella, dau2006, san}). Secondary current sheets may themselves undergo  tearing-type instabilities that lead to the formation of chains of secondary magnetic islands  (or plasmoids). Such a process repeats itself progressively over smaller and smaller spatial and temporal scales via a recursive mechanism of X-point generation, collapse and disruption~\citep{ dau, uzdensky_2010}, that can be described by a geometrical progression~\citep{anna, alkendra}The resulting hierarchy of magnetic islands and current sheets is consistent with a self-similar sequence of fast tearing type instabilities when starting from an ``ideally'' unstable sheet~\citep{anna}, in a way reminiscent of the ``fractal'' reconnection model first introduced  by~\citet{shibata} to  explain energy release in flares.  

Generally, self-similar  processes of energy transfer across scales are associated with energy spectra that follow well defined power-laws, and  we therefore expect that the  ``recursive'' reconnection described above displays a developed energy spectrum as well. Recent work based on particle-in-cell simulations of a collisionless plasma has reported the formation of developed energy spectra within the diffusion region of an unstable current sheet, and it has been shown that the turbulent state generated by secondary magnetic islands  can enhance particle energization~\citep{san}. Modeling and understanding the  mechanisms leading to such an energy cascade inside reconnecting sheets is therefore important for understanding  not only flare dynamics (for which the concept of fractal reconnection was in fact introduced first), but also and more in general for understanding both particle energization and scattering in reconnection events and for its implications in turbulence, where current sheets are an intrinsic and inexorable part of the turbulent cascade. 

Here we focus on the nonlinear stage of the tearing mode, and show that a regime in which magnetic energy density follows a  power-law close to $B(k)^2\sim k^{-0.8}$  exists  within the diffusion region, at the neutral line. We also find that the energy cascade is anisotropic, approximately satisfying $k_\bot\sim k^{3/5}$, where $k_\bot$ represents the inverse of the length scale across the sheet. The energy spectrum changes across the current sheet and farther from the neutral line a Kolmogorov-like spectrum is approached,  $B(k)^2\sim k^{-5/3}$. By analyzing the nonlinear stage of a tearing unstable current sheet and starting from the recursive model discussed in~\cite{anna, anna2}, we propose a quantitative model that explains the observed magnetic energy spectrum. Our result predicts a spectrum power-law and anisotropy in $k$-space which are independent of the current sheet thickness  and that weakly depend on the current sheet configuration.  We finally discuss possible implications for phenomenological theories of turbulence by considering the effect of ``recursive'' tearing on the evolution of intermittent coherent structures. 

\begin{figure}
\begin{center}
\includegraphics[width=0.5\textwidth]{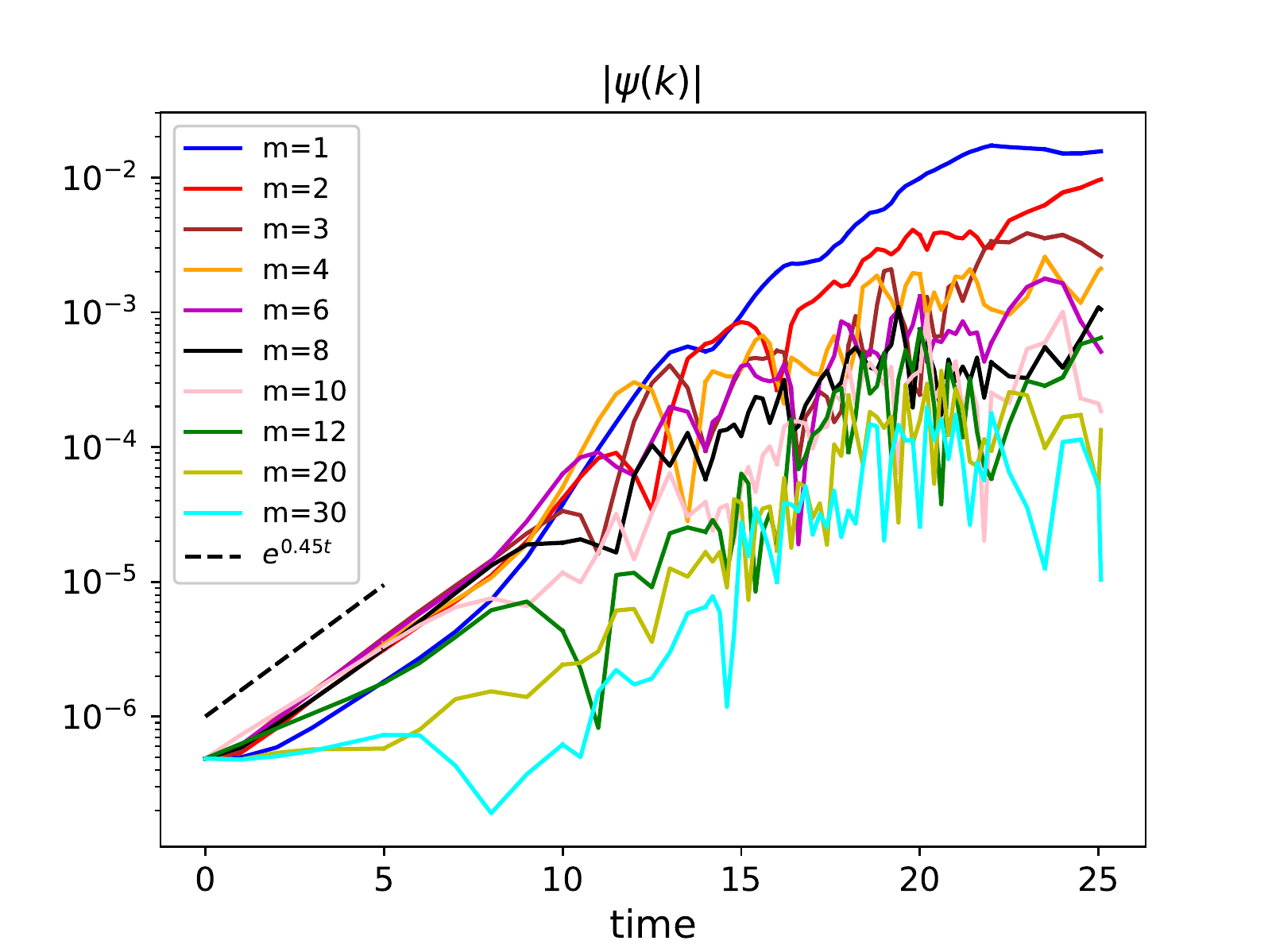}
\caption{\label{modes} Amplitude of some unstable Fourier modes of the flux function $\psi$ as a function of time. The dashed line is shown for reference and it indicates the linear growth rate. }
\end{center}
\end{figure}

\section{Numerical set up and initial conditions}
\begin{figure*}
\begin{center}
\includegraphics[width=0.495\textwidth]{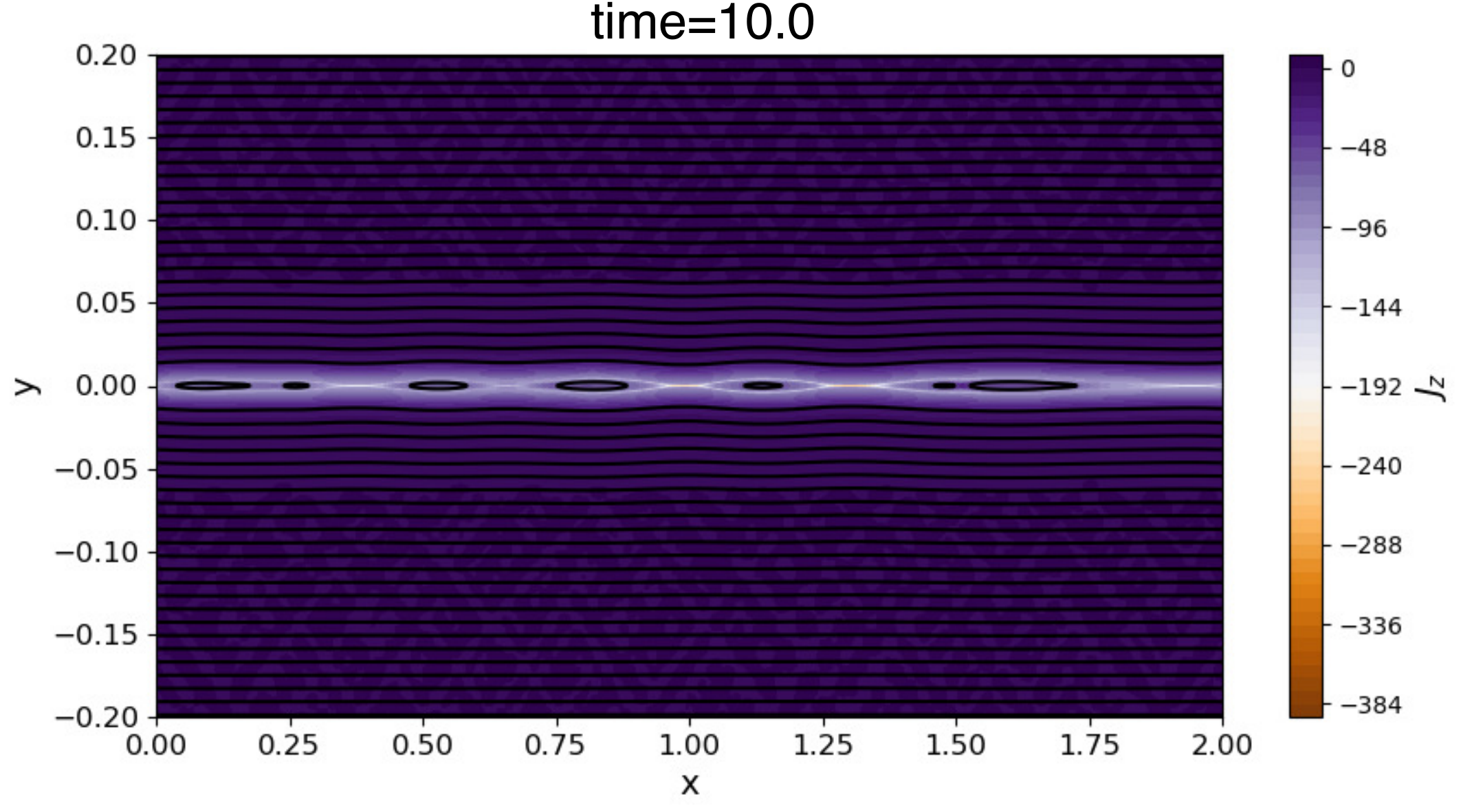}
\includegraphics[width=0.49\textwidth]{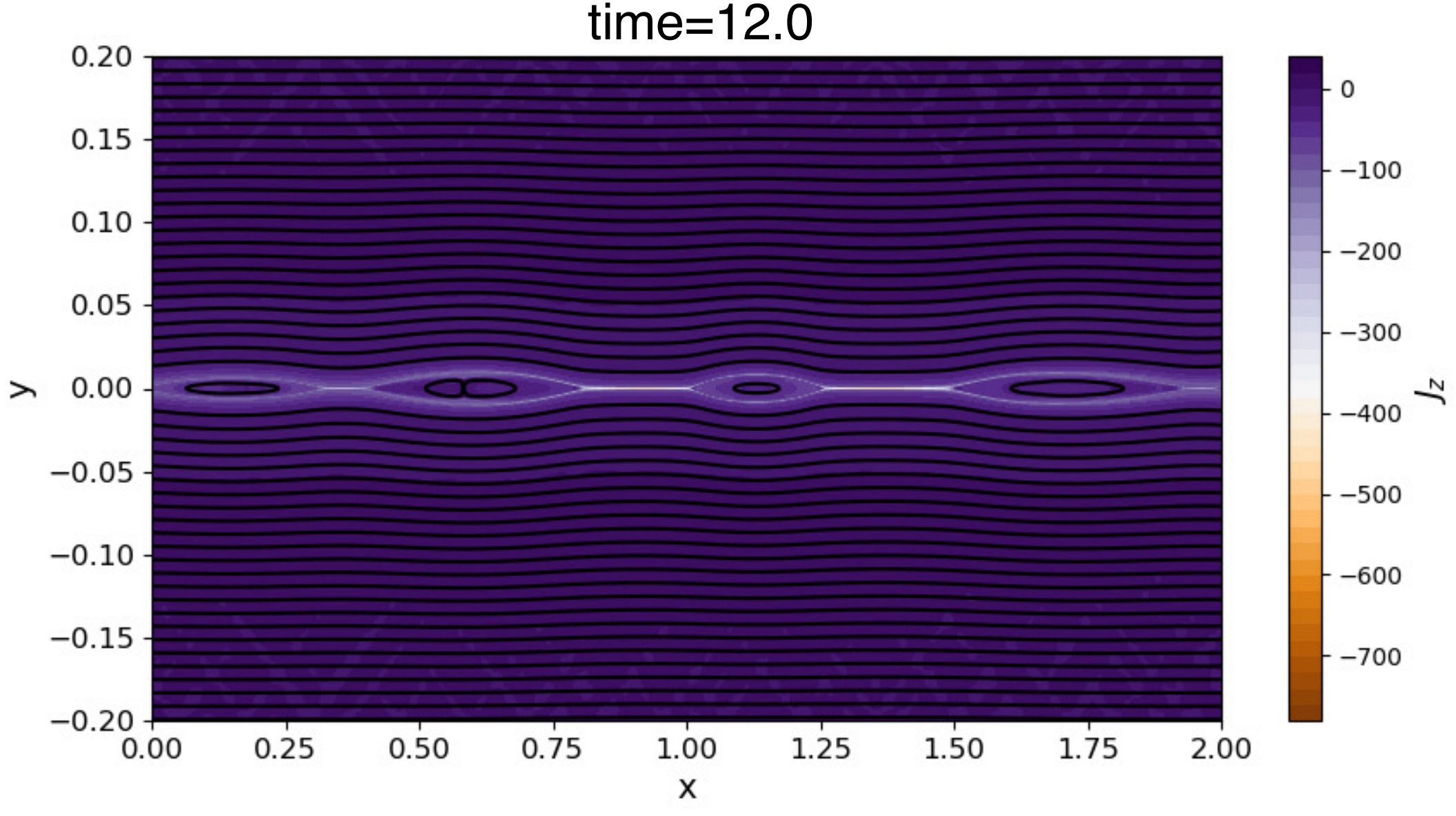}
\includegraphics[width=0.5\textwidth]{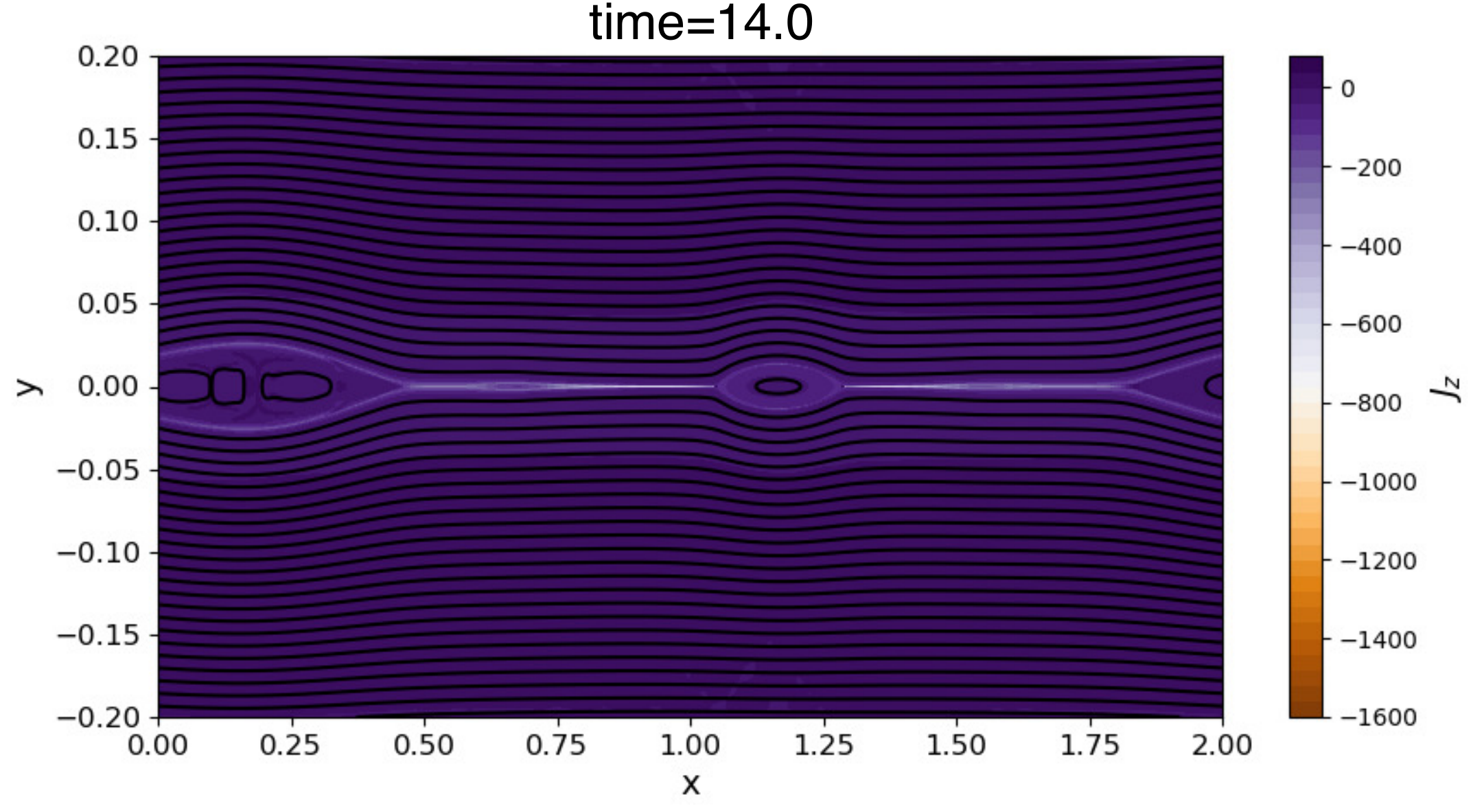}
\includegraphics[width=0.49\textwidth]{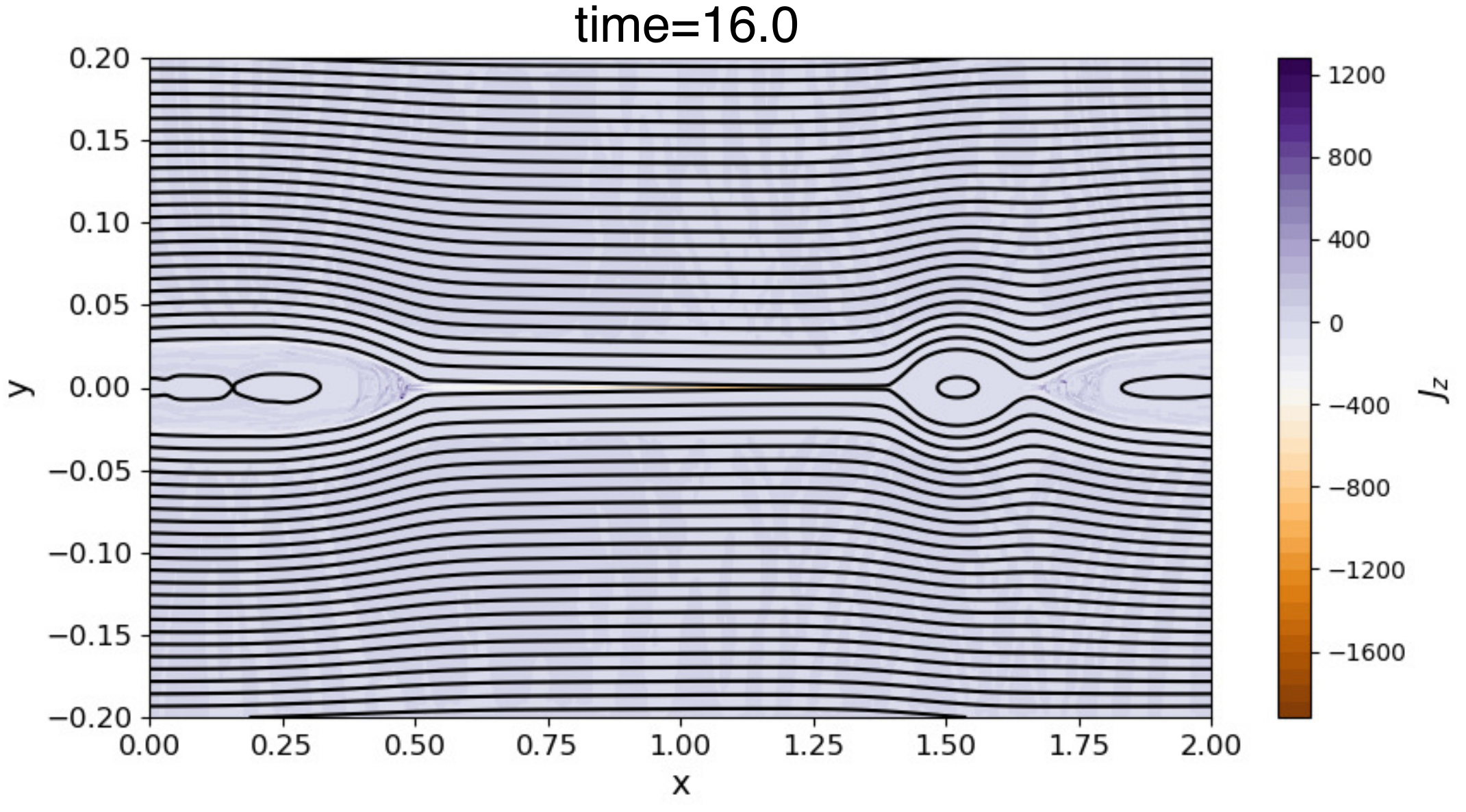}
\includegraphics[width=0.5\textwidth]{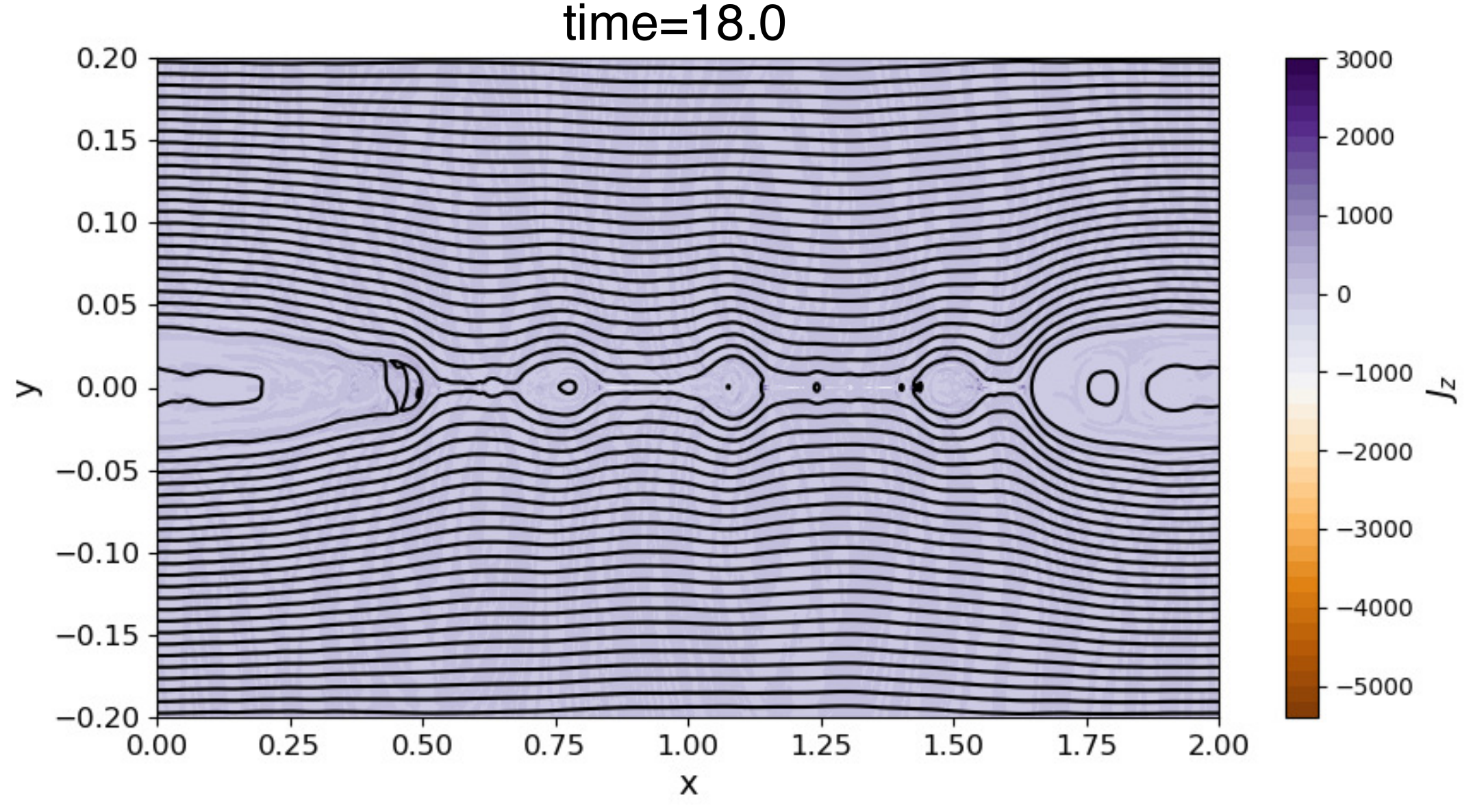}
\includegraphics[width=0.495\textwidth]{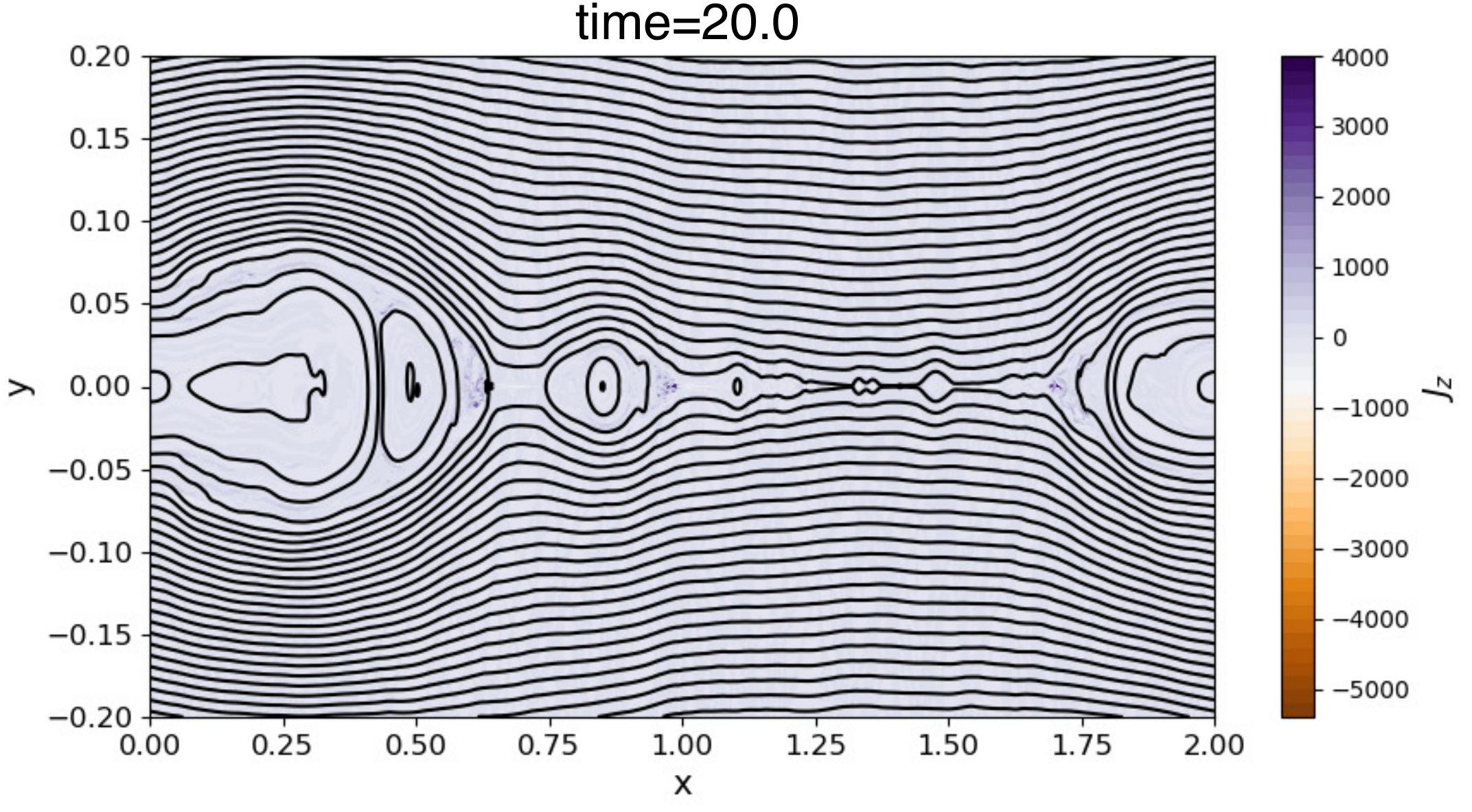}
\includegraphics[width=0.495\textwidth]{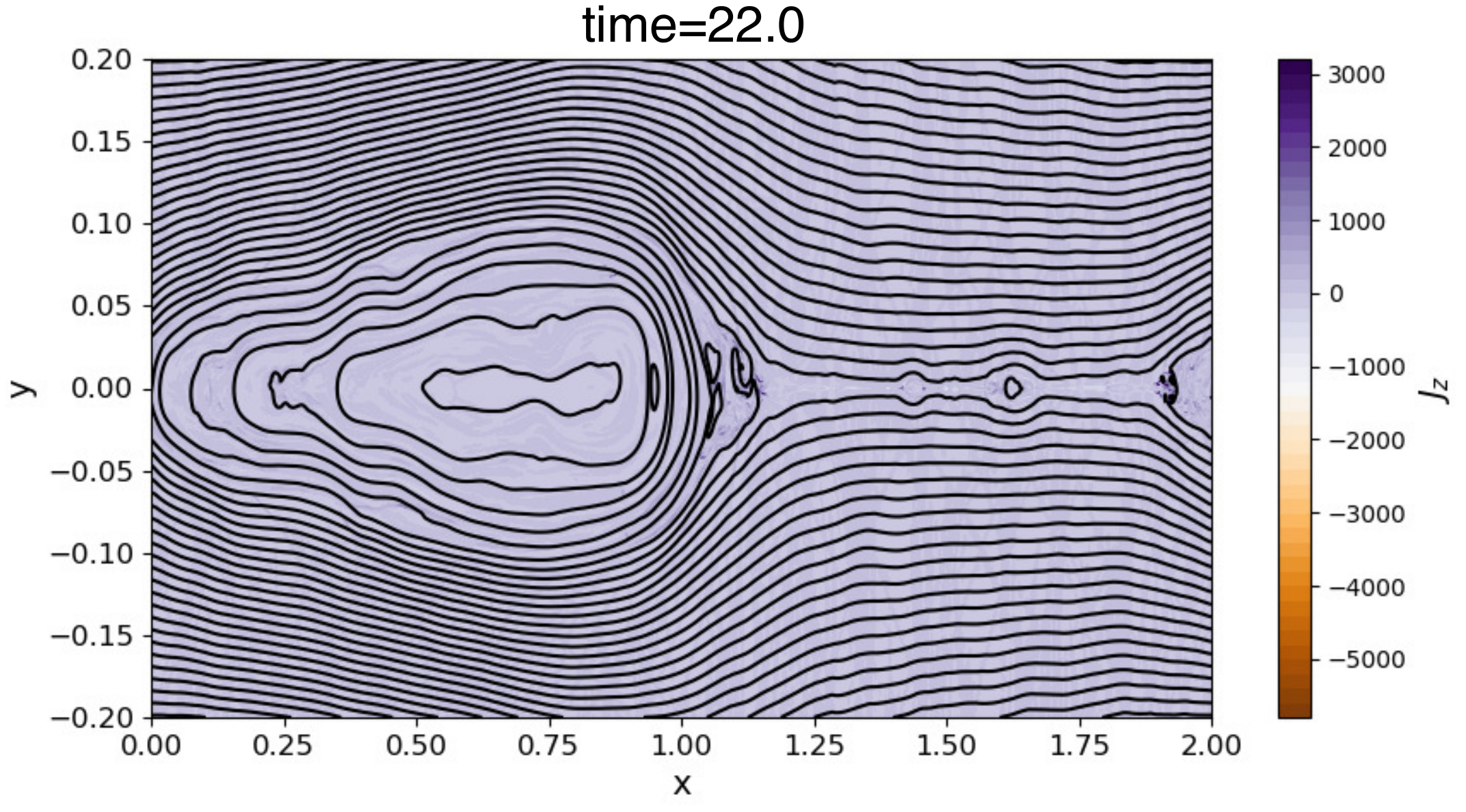}
\includegraphics[width=0.5\textwidth]{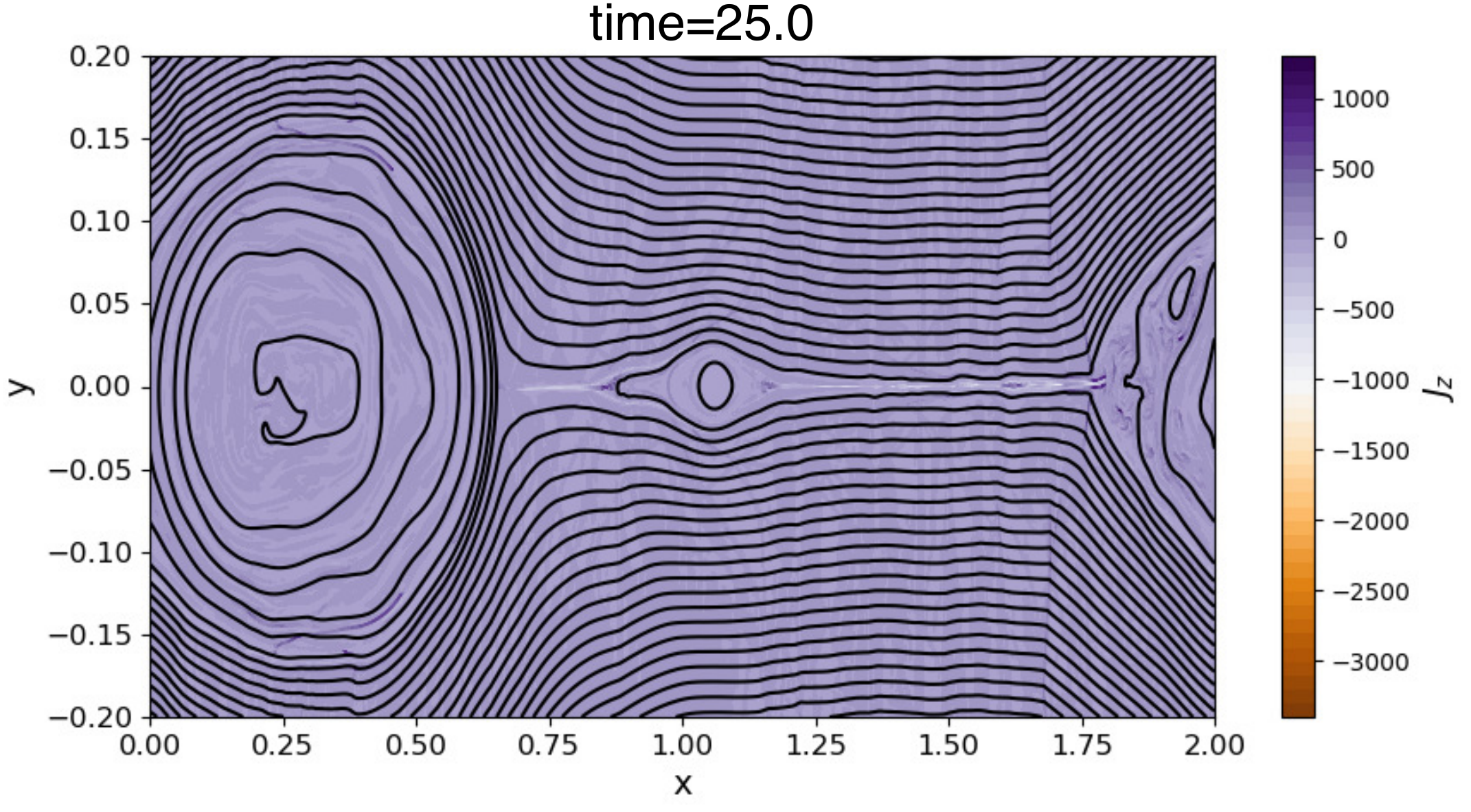}
\caption{\label{contg} Contour plots of the out-of-plane current density $J_z$ (color coded) and of the flux function $\psi$ (black lines) from the early nonlinear stage to $t=25$.  Snapshots are shown, form left to right, top to bottom, at time $t=10,\,12,\,14,\,16,\,18,\,20,\,22,\,25.$}
\end{center}
\end{figure*}

We employ a  2.5D  (two spatial coordinates and three dimensional fields) compressible  MHD code, in which we take an adiabatic closure and a  viscous stress tensor. The code is periodic in the $x$ direction whereas non reflecting boundary conditions are imposed in the inhomogeneous $y$  direction with the method of projected characteristics~\citep{thompson, slandi2005}. Fast Fourier Transform is used along the $x$ direction and a sixth order compact finite difference scheme  is employed along~$y$. A sixth order compact filter is adopted in both $x$ and $y$ coordinate to damp energy accumulation at the grid scale. An explicit fourth order Runge-Kutta method in used for time integration.

As initial condition we choose a plasma equilibrium with homogeneous mass density $\rho_0$ and pressure $p_0$. The  magnetic field is force-free and it is given by a {\color{blue}(force-free)} Harris sheet profile with fixed (half) thickness $a$ plus an out-of-plane magnetic field that ensures magnetic pressure balance: 
\begin{equation}
{\bf B}_0= B_0\tanh\left(  \frac{y}{a} \right){\bf\hat x}+ B_0 \sech\left(  \frac{y}{a} \right) \bf\hat z.
\end{equation}
We normalize the magnetic field  and velocity to $B_0$ and to the Alfv\'en speed $v_a=B_0/\sqrt{4\pi \rho_0}$, respectively,  lengths to the half-length $L$ of the sheet, time to the Alfv\'en time  $\tau_a=L/v_a $, density  and pressure  to the background density $\rho_0$ and to magnetic pressure $B_0^2/4\pi$, respectively. As usual, it is useful to introduce also the flux function $\psi$ such that the in-plane magnetic field is ${\bf B_\bot=\nabla\times\psi{\bf \hat z}}$. In this simulation the magnetic diffusivity is set to $\eta=10^{-6}$, so that the macroscopic Lundquist number defined with the half-length of the current sheet is $S\equiv Lv_a/\eta=10^6$, the Prandtl number $P$ is set to one, the equilibrium pressure $p_0=0.8$,  and we choose a current sheet aspect ratio $L/a=S^{1/3}=100$. The simulation box has sides $L_x\times L_y=(2\times 0.4)L$  with $4096^2$ mesh points.  We seed the instability with a random noise of magnetic fluctuations with rms amplitude $\delta b\simeq0.01$ and wave numbers $m$ in the periodic $x$-direction in the interval $1\leq m\leq1024$,  and localized along $y$ within the magnetic field shear region.

\section{Numerical results}
\subsection{Overview of the linear and nonlinear stages}
In Fig.~\ref{modes} we show for reference the temporal evolution of some unstable Fourier modes of $\psi$ taken along the neutral line $y=0$. During the initial phase of the simulation, from $t=0$ until about $t\simeq9$, we see the exponential growth of several modes close to the most unstable one $m=3$,  corresponding to wave vector $k=m2\pi /L_x\simeq10$. The inferred growth rate is about $\gamma\simeq 0.45$ (dashed line in Fig.~\ref{modes}), slightly smaller than the one expected theoretically with the present parameters ($\gamma=0.47$) as compressibility tends to decrease reconnection rates. At the end of the linear stage, towards $t=9$, a superposition of modes from  $m=4$ to $m=6$ has grown. At that time the nonlinear stage begins, corresponding to a faster growth of low wave number modes (essentially $m=1-4$) and to the emergence of two intense  currents. In general the nonlinear stage is characterized by a competition and interplay between  coalescence of magnetic islands (inverse cascade) and X-point collapse and disruption into smaller scale secondary sheets  in correspondence of the most intense currents (direct cascade). We give an overview of the overall evolution of the current sheet from the early nonlinear stage until saturation of the inverse cascade in Fig.~\ref{contg}.  

Fig.~\ref{contg}  displays a set of contour plots of the out-of-plane current density $J_z$ (color coded) and of $\psi$ (black lines). During the early nonlinear stage (between $t\simeq9$ and $t\simeq14$), merging of magnetic islands and X-point collapse lead to the formation of two islands separated by  two secondary current sheets near $x=1$ and $x=1.5$, which both become unstable to tearing at about $t=14$ (see Fig.~\ref{contg}, first and second rows). Subsequently, magnetic island merging leads to roughly one elongated magnetic island with a single, highly inhomogeneous current sheet (thicker on the left end side than on the right end side,  Fig.~\ref{contg}, second row second column), within which  a sequence of tearing type instabilities that generate smaller and smaller structures (Fig.~\ref{contg}, second last and last columns) is triggered in a recursive fashion as reported in previous work~(e.g., \cite{dau, anna}).  As can be seen, the overall nonlinear dynamics  leads to a final state characterized by one big magnetic island, when the mode $m=1$ has reached a steady amplitude level (cfr. also Fig.~\ref{modes}), and an elongated current sheet with embedded small scale secondary islands and sheets.
\begin{figure}
\begin{center}
{\includegraphics[width=0.515\textwidth]{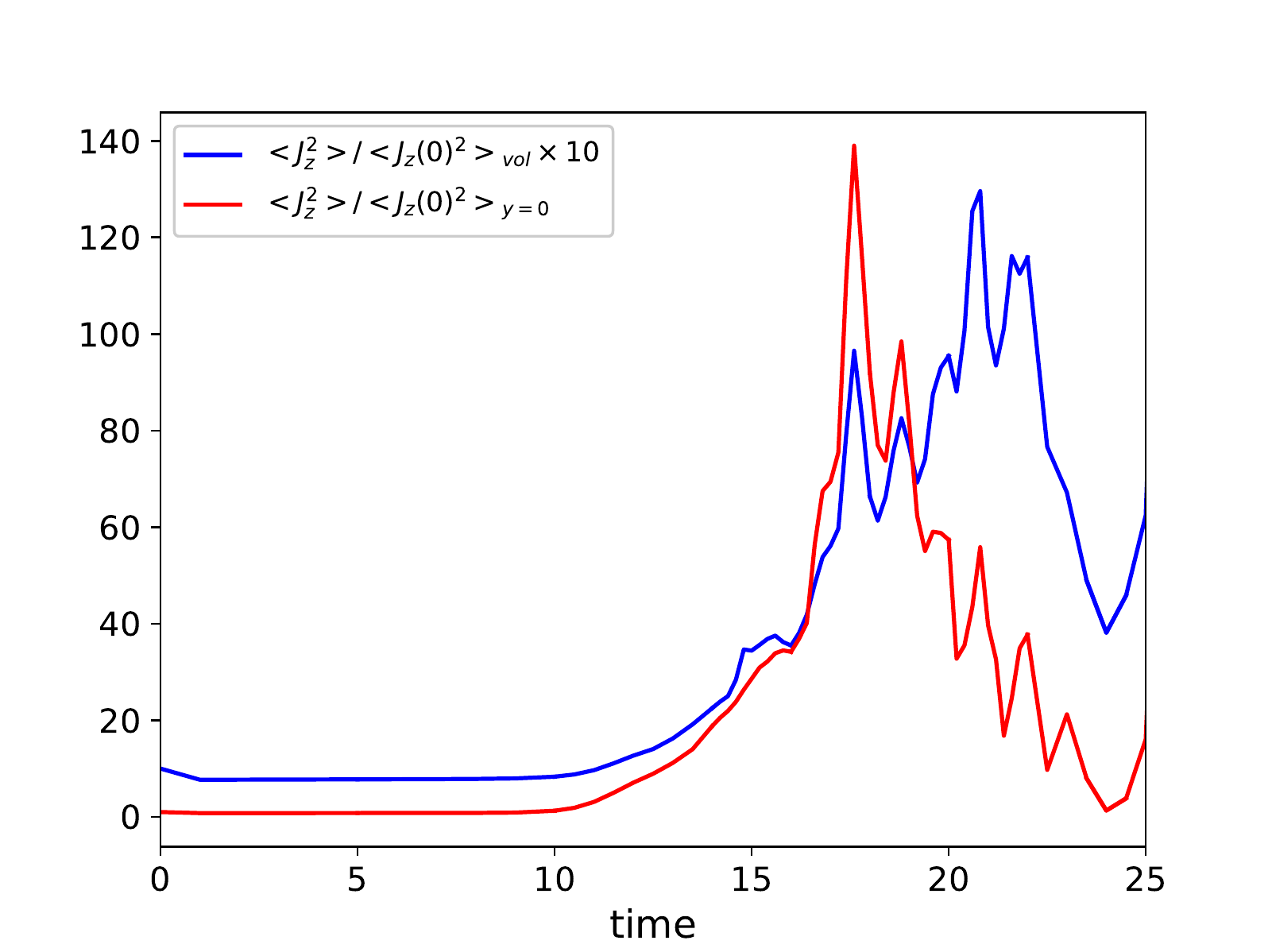}}
{\includegraphics[width=0.475\textwidth]{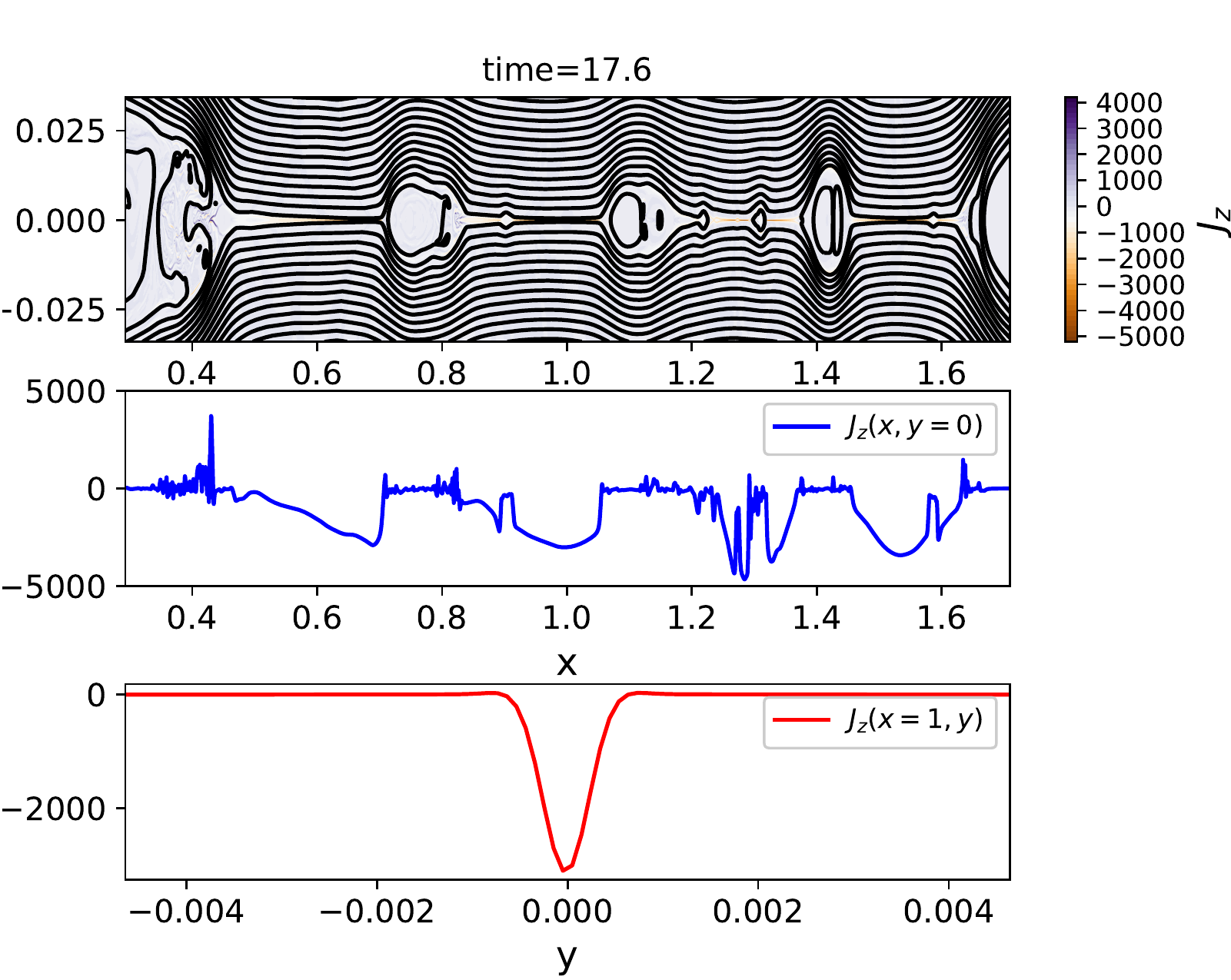}}
\caption{\label{current} Top: time evolution of the normalized average current density $<J_z^2>$  over the whole simulation box (blue line, multiplied by a factor of 10 to enhance visibility) and along the neutral line (red color). Bottom: blow out of the unstable secondary current sheet at $t=17.6$. Upper panel: contour plot of $J_z$ (color coded) and of $\psi$ (black lines). Middle panel: cut of $J_z$ as a function of $x$ at the neutral line ($y=0$). Bottom panel: cut of $J_z$ across the current sheet at $x=1.0$.}
\end{center}
\end{figure}

\subsection{Current density evolution}
The different stages  in the overall evolution of the current sheet described in the previous section can be recognized also in Fig.~\ref{current}, left panel, where we show  $< J_z^2 >$ as a function of time,  the brackets denoting spatial average: the blue color corresponds to the average over the simulation box (and multiplied by a factor of 10 to enhance visibility), while the red color corresponds to the average at the neutral line ($y=0$), i.e., inside the diffusion region; both quantities are normalized to their initial values.  The evolution of $<J_z^2>_{y=0}$ at the neutral line (red color) shows an initial quiet stage, followed by a rapid rise and a subsequent fall towards smaller average values with superposed impulsive events  (the spikes).   The rising phase begins with an almost monotonic increase of $<J_z^2>_{y=0}$,  corresponding to the formation of the first two secondary currents (visible between $t=10$ and $t=14$ in the plot) and the formation of the second elongated current sheet before it becomes unstable again at about  $t= 16.4$. In that time interval, the average current growth slows down around $t=16$ because of the interplay between the formation of the second current sheet and the merging of two magnetic islands (cfr. Fig.~\ref{contg}, second row, second column).   After that time there is a nice sequence of spikes, with each spike corresponding to the formation of new, smaller and more intense current sheets (increase phase of the spike), and their relaxation via tearing instability (decreasing phase of the spike). The  profile of $<J_z^2>_{y=0}$ has  a maximum at $t\simeq 17.6$. For reference, we show  in Fig.~\ref{current}, bottom panel, a blow out of the secondary current sheet at $t=17.6$ that displays the contours of $J_z$ and $\psi$ (upper plot),  a cut of the current density along the sheet, $J_z(x,y=0)$ (middle plot) and a cut across the sheet at $x=1$, $J_z(x=1,y)$  (lower plot). We remark that the small scale, large amplitude oscillations seen in the middle right panel of Fig.~\ref{current} are well resolved and do not correspond to the well known Gibbs numerical effect arising when structures are poorly resolved.

As can be seen from Fig.~\ref{current}, bottom panels, and in general by inspection of Fig.~\ref{contg}, the most intense currents are localized at the neutral line where $J_z(x,0)$ evolves into intense filaments with local values reaching up to $50$ times the initial current density. The profile of $<J_z^2>$ averaged over the entire box (blue color) displays the same spikes although the average current has a longer rising phase, as it increases until about $t=21$. This is due to the fact that current density spreads farther from the neutral line due to the growing size of magnetic islands during merging, and in fact $<J^2>$ begins to decrease when the mode $m=1$ has reached a steady value and  the inverse cascade has  saturated. At the very end of our simulation the current starts to increase again, and we ascribe this effect not to reconnection but to the development of sharp gradients at the edges of the main magnetic island, clearly visible  in the vicinity of $x=1.75$ in the very last panel of Fig.~\ref{contg}.

\begin{figure}
\begin{center}
\includegraphics[width=0.5\textwidth]{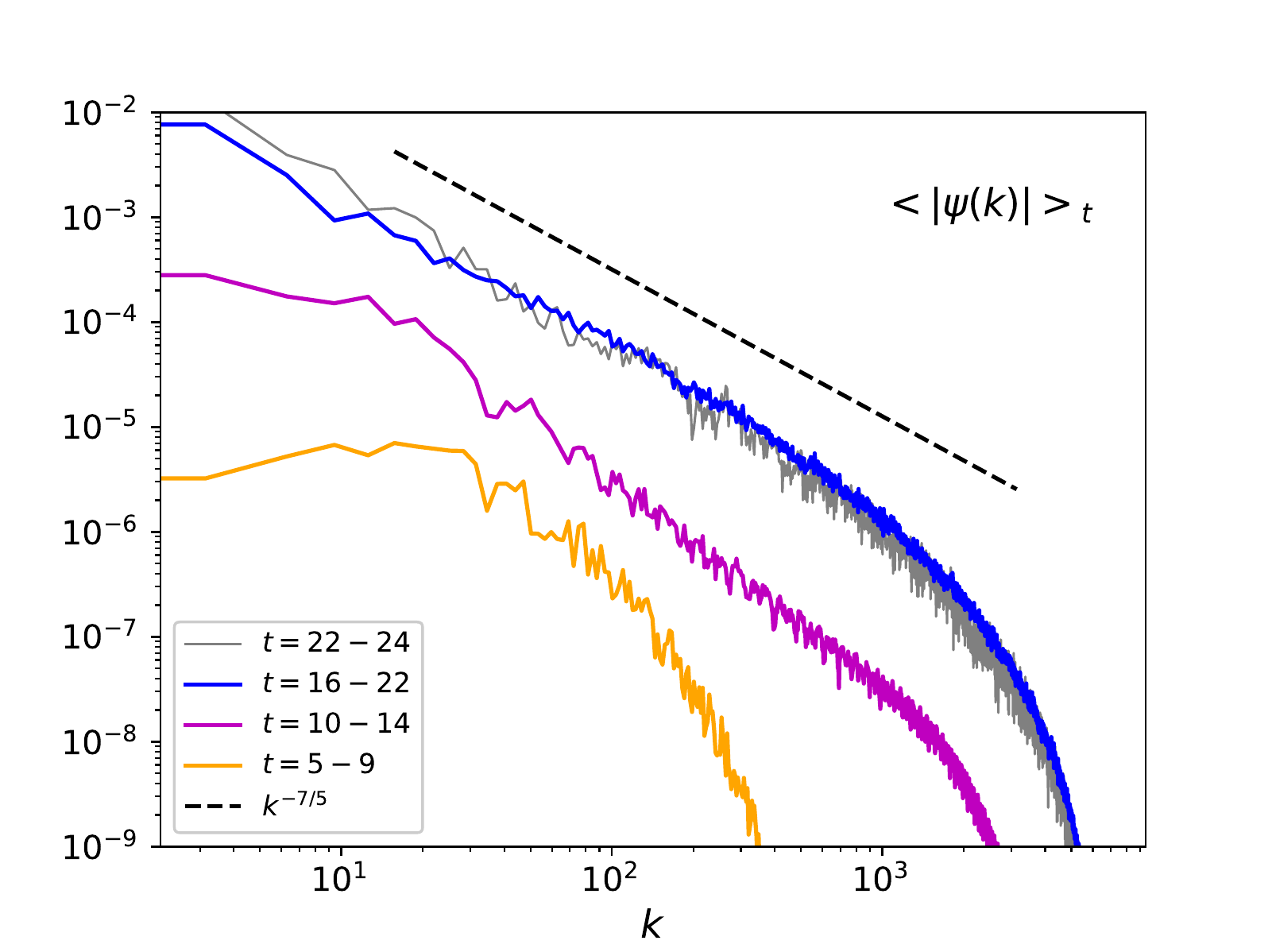}
\caption{\label{spe} Amplitude of  $|\psi(k)|$ at $y=0$ during three different stages, averaged over the time intervals indicated in the legend. The orange color corresponds to the linear stage, the magenta color to the early nonlinear stage, the blue and grey color correspond to the time interval with ``recursive'' reconnection and multiple tearing type instabilities.}
\end{center}
\end{figure}

\subsection{Energy spectra}

We consider $t=16$ the time at which ``recursive'' reconnection begins to be fully at play.  In Fig.~\ref{spe} we show the amplitude of $\psi$ in Fourier space at three different stages of the evolution. The spectrum is taken along the  $x$-direction at the center of the sheet ($y=0$) where magnetic islands are generated, and we have averaged $|\psi(k)|$ over the time interval indicated in the plot legend. The blue and grey curves of $|\psi(k)|$ in Fig.~\ref{spe} show that a steady-state  spectrum is formed during the ``recursive'' reconnection: the slope of the spectrum appears to remain only slightly flatter than a simple power law $\psi(k)\sim k^{-7/5}$ (black dashed line) which is approached at wave vectors $k\gtrsim15$ (close to the dominant mode at the end of the linear stage) until about $k\simeq 300$, after which the spectrum steepens.  Since the slope is  sensitive to the lower and upper bound of the wave vector range considered, we performed several numerical fits around the range $\Delta k=(15,\,300)-(10,400)$, and inferred a slope $\alpha_k=-1.39\pm0.03$.
\begin{figure}
\begin{center}
\includegraphics[width=0.5\textwidth]{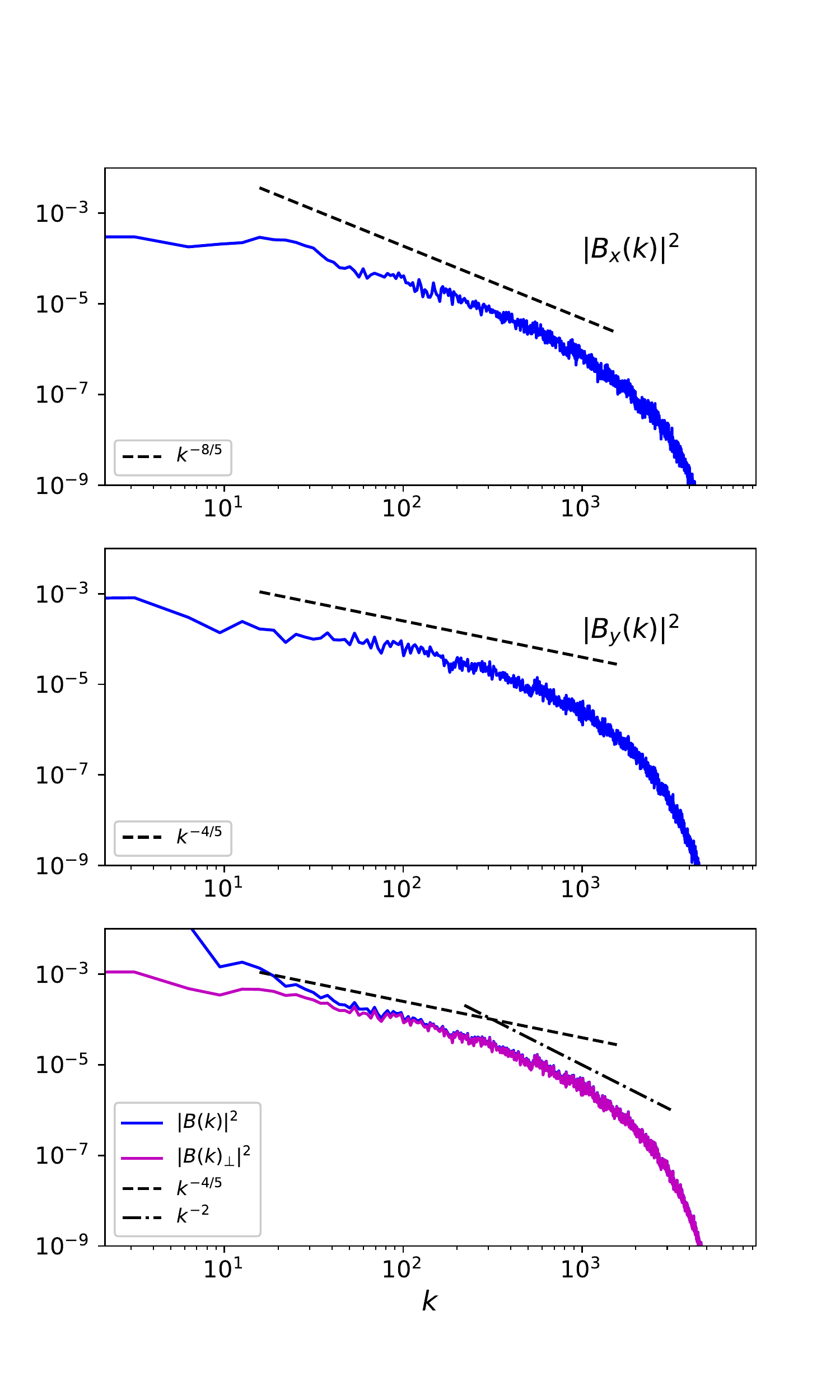}
\caption{\label{spectrum_B_0} Magnetic energy density spectra at the neutral line $y=0$, obtained after averaging spectra at times   $t=16-22$. Upper panel: $|B_x(k)|^2$. Middle panel:  $|B_y(k)|^2$ (the reconnecting magnetic field). Bottom panel:  total $|B(k)|^2$ (blue) and in-plane $|B(k)_\bot|^2$ (magenta) magnetic energy density spectra.}
\end{center}
\end{figure}

\begin{figure}
\begin{center}
\includegraphics[width=0.5\textwidth]{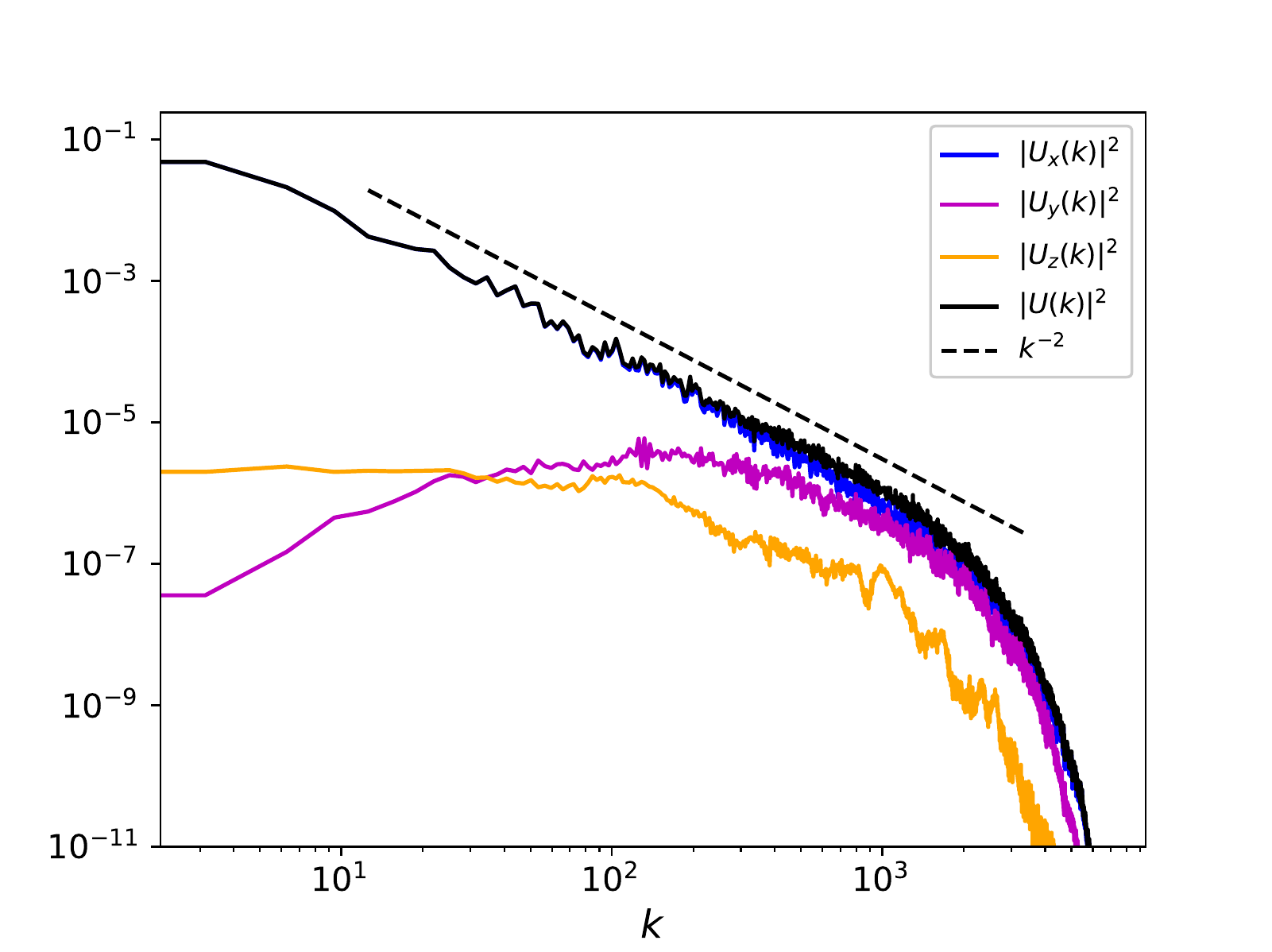}
\caption{\label{spectrum_U_0} Kinetic energy density spectra at the neutral line $y=0$, obtained after averaging spectra over times $t=16-22$.  Velocity field components are in colored lines, the total kinetic energy in black.}
\end{center}
\end{figure}

\begin{figure}
\begin{center}
\includegraphics[width=0.49\textwidth]{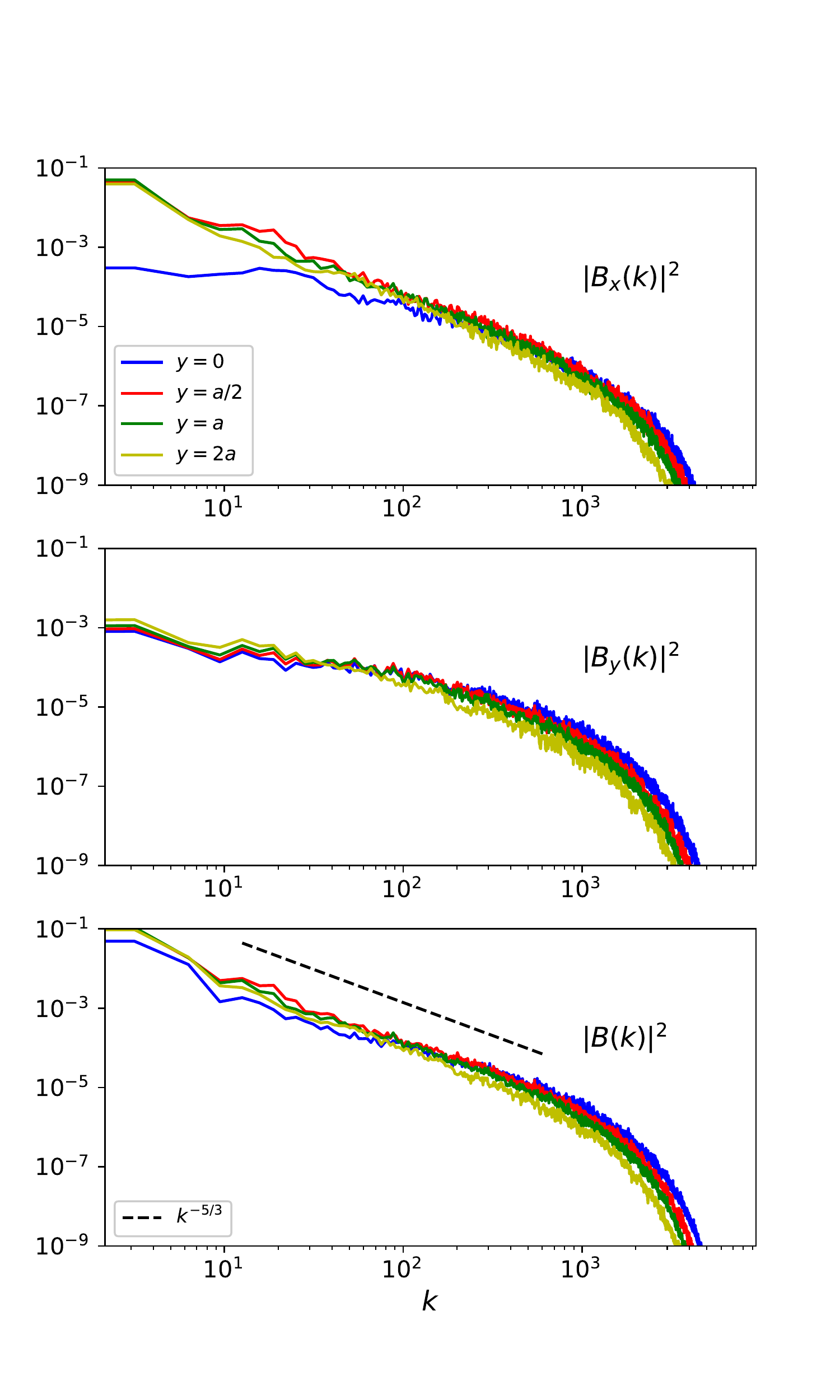}
\caption{\label{spe_sheetB} Magnetic (left panels) and kinetic (right panels) energy density spectra along the sheet at various values of $y$ as indicated in the legend. The first two rows show in-plane energy density and the last row the total energy density.}
\end{center}
\end{figure}

\begin{figure}
\begin{center}
\includegraphics[width=0.49\textwidth]{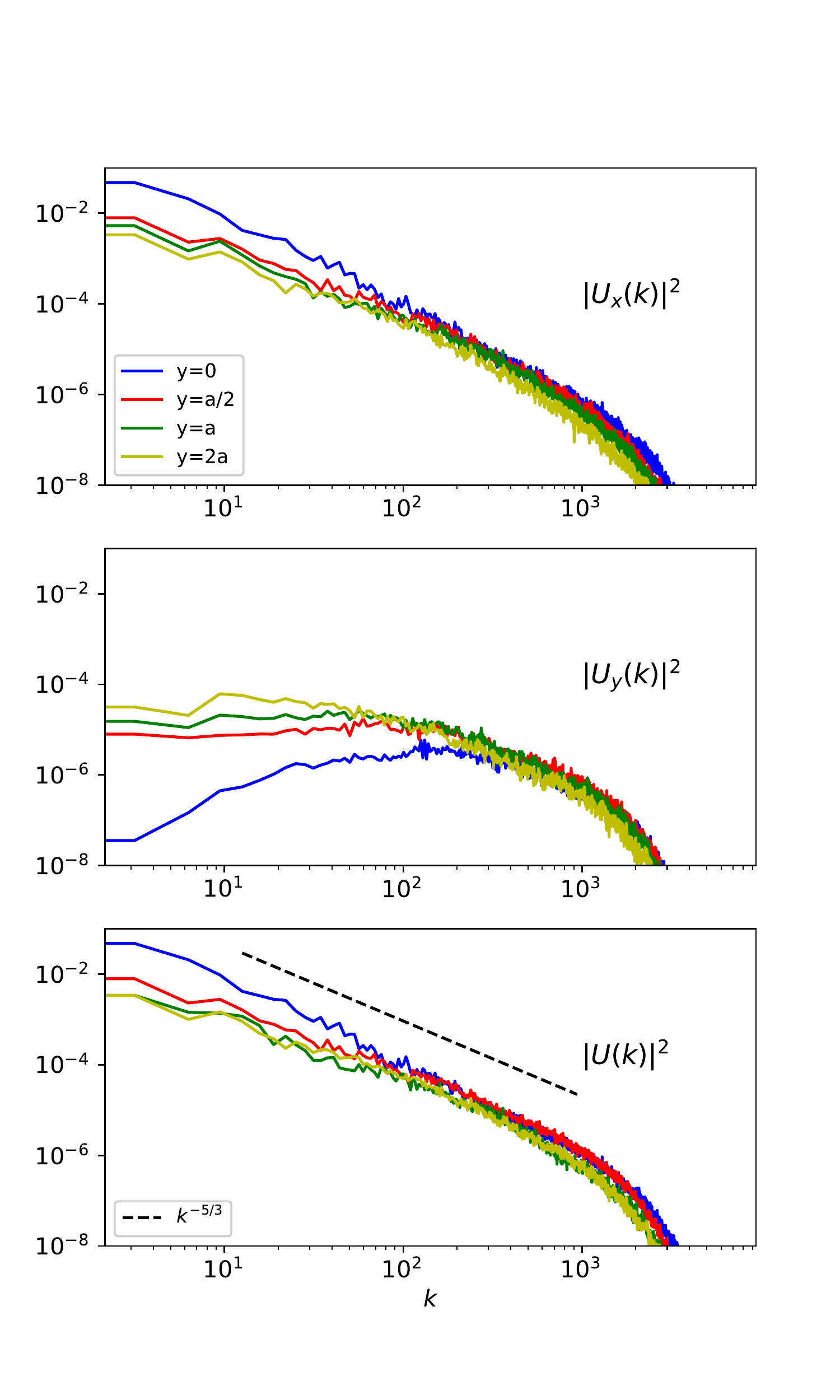}
\caption{\label{spe_sheetU} Magnetic (left panels) and kinetic (right panels) energy density spectra along the sheet at various values of $y$ as indicated in the legend. The first two rows show in-plane energy density and the last row the total energy density.}
\end{center}
\end{figure}

In Fig.~\ref{spectrum_B_0} and Fig.~\ref{spectrum_U_0} we show the magnetic and kinetic energy density spectra along the neutral line, at $y=0$. Energy spectra are displayed in solid black and colored lines and, for reference, we also show some  spectral slopes in dashed and dot-dashed lines, to be discussed later in section~\ref{discussion}. As shown in Fig.~\ref{spectrum_B_0}, we find that  $|B_x(k)|^2\sim k^{-1.24\pm0.06}$ (estimated according to  the same criterium adopted for $\psi(k)$), a bit flatter than the shown scaling $k^{-1.6}$ (upper panel) while the   reconnecting component of the magnetic field scales to a good approximation as $|B_y(k)|^2\sim k^{-0.79\pm0.07}$ (middle panel). The  $B_z$ component (not shown) dominates only at the lower wave vectors (about $k\lesssim 15$) while at larger wave vectors energy is determined by the in-plane magnetic field (bottom panel), displaying an overall power law scaling $k^{-0.98\pm0.06}$. At higher wave vectors, $k\gtrsim 300$, a slope close to $k^{-2}$ is developed ($ k^{-1.9\pm0.1}$), ascribed to the development of shocks at the edges of magnetic islands. As can be seen from Fig.~\ref{spectrum_U_0},  strong flows along the neutral line naturally develop as a consequence of reconnection, where $U_x$ is the dominant component by orders of magnitude. Flows are dominated by shocks at the magnetic island leading edges, consistent with a spectrum $|U_x(k)|^2\sim k^{-1.95\pm0.1}$. Each spectrum  evolves across the current sheet. This is shown in Fig.~\ref{spe_sheetB} and Fig.~\ref{spe_sheetU}, which display magnetic  and kinetic  energy spectra, respectively,  at different values of  $y$ as indicated in the legend. While moving outwards farther from the neutral line, the magnetic  energy  becomes steeper and it approaches a Kolmogorov-like distribution with spectral slope close to $-5/3$  (we estimated spectral slope $\alpha_k=1.72\pm0.01$). Kinetic energy displays an opposite trend, with decreasing energy outside the sheet (strong flows are localized at the neutral line) but  reaching a slope of $-5/3$ similarly to the magnetic field   ($\alpha_k=1.51\pm0.06$); for $k\gtrsim 300$,  the slope becomes steeper,  close to $-3$ (not shown).  

\begin{figure}
\begin{center}
\includegraphics[width=0.5\textwidth]{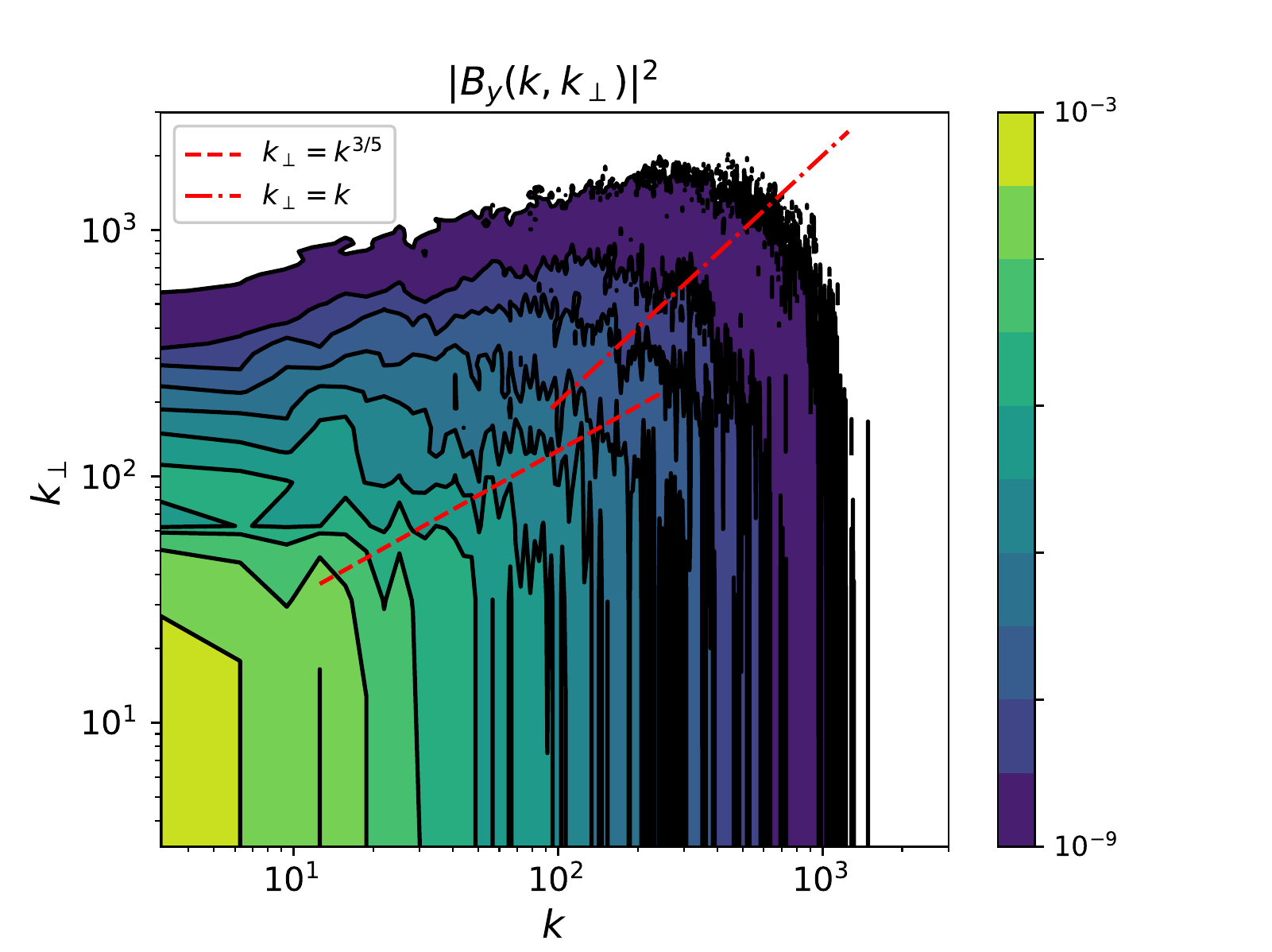}
\caption{\label{spectrum_2D}Two-dimensional spectrum of the reconnecting magnetic field averaged over times $t=18-20$.}
\end{center}
\end{figure}

Finally, Fig.~\ref{spectrum_2D} shows the 2D spectrum of $B_y$. There is clear evidence for an anisotropic cascade at intermediate values of $(k, \, k_\bot)$, where $k_\bot$ indicates the wave vector along $y$; red lines represent indicative trends of $k_\bot$ vs $k$. In order to perform the 2D Fourier transform we have applied a Hann window to reduce possible boundary effects along the $y$-direction, although results are very similar to what was obtained without windowing.


\section{Discussion}
\subsection{Transition to turbulence in a single current sheet}
\label{discussion}
We adopt here a heuristic method to derive a model that predicts specific scaling laws of magnetic energy spectra, following the idea that magnetic reconnection provides a channel for reaching  small   scales via a scale-invariant (recursive) hierarchy of tearing modes. To this aim, we consider a simplified model by assuming an inviscid, incompressible plasma and by neglecting the effect of flows on the stability of current sheets.  

On the basis of similarity and rescaling of quantities, it is convenient now to label with an $n$ the typical length scales that describe the  current sheet generated at step $n$ of the hierarchy, with $n\geq 1$: the half length $L_n$  and thickness  $a_n$, the inner singular layer half thickness $\delta_n$, and the wave vector of the most unstable mode $k_n$.  The local Lundquist number is defined accordingly as $S_n=L_n v_a/\eta=(L_n/L)\, S$, where $L$ and $S$ refer to the primary current sheet. We therefore have assumed that as long as the recursive process of reconnection takes place, a given current sheet at step $n$ has a sufficiently large aspect ratio so as to allow for the fastest mode to develop. Also, for the sake of simplicity, we assume that the upstream Alfv\'en speed does not change significantly. In this regard we have verified that the upstream magnetic field of the secondary sheets is about $80\%$ of $B_0$ and we therefore neglect embedding effects~\citep{cassak_2009, delsarto_2017} by normalizing speeds to $v_a$. 
 
 We recall briefly the properties of the most unstable mode in thin current sheets, and in particular the scalings with $S_n$ of the wave vector and  inner singular layer of the most unstable mode: 
 \begin{equation}
k_na_n\sim S_n^{-1/4}\left(\frac{a_n}{L_n}\right)^{-1/4},
\label{fgm0}
\end{equation}
 \begin{equation}
 \frac{\delta_n}{a_n}\sim S_n^{-1/4}\left(\frac{a_n}{L_n}\right)^{-1/4}.
\label{fgm2}
\end{equation}

We estimate the nonlinear amplitude of the flux function $\psi_n$, which is associated to wave vector $k_n$,  from the condition $w_n\simeq\delta_n$, where $w_n$ is the magnetic island width, $\psi_n/(a_n B_0)\sim (w_n/a_n)^2$ (see, e.g., \cite{bisk2000}). This condition, typically used to estimate when nonlinear effects cannot be neglected during magnetic island growth, yields the following scaling: 
 \begin{equation}
\frac{ \psi_n}{B_0 L_n}\sim 0.25\,S_n^{-1/2}\,\left(\frac{a_n}{L_n}\right)^{1/2}.
\label{psi}
\end{equation}

Equation~(\ref{psi}) can be used to find the relation between $\psi_n$ and the length scale $k_n$ at which magnetic energy is conveyed.  We now consider the more general case in which we assume that a tearing instability at step $n$ is triggered when a generic aspect ratio  $a_n/L_n\sim S_n^{-\alpha}$ is reached (for us it should be, in principle, $\alpha=1/3$). With this assumption we obtain  
\begin{equation}
{\psi_n}\sim \left(\frac{L_n}{L}\right)^{-\frac{1}{2}(\alpha-1)}\,S^{-\frac{1}{2}(\alpha+1)}.
\label{psi_n}
\end{equation}

 Since the fastest growing mode within $L_n$ is $k_nL_n\sim S_n^{-1/4}(a_n/L_n)^{-5/4}$, we can write
 \begin{equation}
\frac{L_n}{L}\sim (k_n L)^{\frac{4}{5}\frac{1}{(\alpha-1)}}S^{-\frac{1}{5}\frac{5\alpha-1}{(\alpha-1)}}.
\label{L_k}
\end{equation}

At this point we can use eq.~(\ref{L_k}) into eq.~(\ref{psi_n}) to express the nonlinear amplitude of $\psi_n$ in terms of $k_n$. This latest step eliminates the parametric dependence on~$\alpha$. We finally go to the continuum limit, as we are interested in the spectral density, obtaining in this way 
\begin{equation}
{\psi(k)}\sim  S^{-3/5}\, (kL)^{-7/5}. 
\label{spectral}
\end{equation}

It is interesting to note that the reasoning followed to find a scaling for $\psi(k)$ with $k$ leads to a universal value of the spectral index, independent of the thickness  of the current sheets, i.e., our result does not depend on $\alpha$.  However, simulations with different initial thicknesses, and at larger values of $S$, are necessary to verify whether and in which way such spectral signatures vary.  The slightly flatter slope observed in Fig.~\ref{spe} might be due to  effects  that have been neglected in our simple model, including  the fact that ``recursive'' reconnection does not take place homogeneously in space at the neutral line and coalescence of the smaller magnetic islands with the larger ones, which could contribute to a redistribution of energy even at $k>15$ (approximately the dominant mode at the end of the linear stage). However, the fact that the spectrum is  very close to a power law with exponent $-7/5$  suggests that the recursive process provides a good description of the nonlinear interactions leading to the direct energy cascade. The recursive reconnection is expected to develop until the local Lundquist number reaches the critical value $S_c$, and current sheets are  stabilized. 

The spectrum of the reconnecting magnetic field  component can be derived in a  straightforward way, because $B_y(k)=k\psi(k)$, which leads directly to $B_y(k)^2\sim k^{-4/5}$, in good agreement with our numerical results. In order to estimate the spectrum of $B_x$ instead, we have  to find a relation between spatial scales along ($k^{-1}$) and across ($\delta$) the sheet. To this aim we combine eq.~(\ref{fgm0})--(\ref{fgm2})  with eq.~(\ref{L_k}), obtaining  $k_n\delta_n\sim S^{-2/5} (k_nL)^{2/5}$, that yields $\delta/L\sim S^{-2/5}  (kL)^{-3/5}$ or also $k_\bot L\sim S^{2/5}(kL)^{3/5}$, not far from what is shown in Fig.~\ref{spectrum_2D}. We finally employ the condition $\boldsymbol\nabla\cdot {\bf B}=0$,  $kB_x(k)\sim B_y(k)/\delta$, to get $B_x(k)^2\sim k^{-8/5}$, a bit steeper than what is observed in the simulation. 

We conclude this section by noticing that there might be circumstances in which current sheets may not  be well described by a Harris-type profile, such as current sheets generated via a turbulent cascade. It is therefore of interest to see how the power-laws change by assuming a generic current density profile. Dependence on the type of  profile enters via the $\Delta^\prime$ parameter~\citep{FKR}, which can be taken as scaling with a generic negative power $p$ of the wave vector, $\Delta^\prime\sim (ka)^{-p}$, where $p=1$ corresponds to the Harris sheet while for instance $p=2$ corresponds to a magnetic field that goes to zero at the boundaries~\citep{delsarto_2016, pucci_2018}. For a generic current density profile we can therefore  generalize our scaling laws as follows (see also~\cite{alkendra}): 
\begin{equation}
 k_n a_n\sim S_n^{-\frac{1}{1+3p}}\left(\frac{a_n}{L_n} \right)^{-\frac{1}{1+3p}},
\end{equation}
\begin{equation}
\frac{\delta_n}{a_n}\sim S_n^{-\frac{p}{1+3p}}\left(\frac{a_n}{L_n} \right)^{-\frac{p}{1+3p}}.
\end{equation}

By repeating the same  exercise as for the Harris sheet, but with these new scalings, we obtain a $\psi(k)$ and anisotropy in $k$-space which are again independent of the thickness, but that are now parameterized in $p$,
\begin{equation}
\psi(k)\sim S^{-\frac{(1+2p)}{(2 + 3 p)}}(kL)^{-\frac{3+4p}{2+3p}}\quad k_\bot L\sim S^{\frac{1+p}{2+3p}}(kL)^{\frac{(1+2p)}{(2+3p)}}.
\label{scaling_p}
\end{equation}

In summary, if $p$ could take any value in the interval $1\leq p<\infty$, we would find that the reconnecting magnetic field  and the anisotropy follow  power-law scalings bounded between two rather close values,
\begin{equation}
B_y(k)^2\sim (kL)^{-4/5}, \quad  k_\bot L\sim (kL)^{\frac{3}{5}}, \quad {p=1},
\end{equation}
\begin{equation}
B_y(k)^2\sim (kL)^{-2/3}, \quad  k_\bot L\sim (kL)^{\frac{2}{3}} , \quad {p\rightarrow\infty}.
\end{equation}

\subsection{Implications for turbulence phenomenology} 

Because the recursive collapse leads to reconnection occurring on time-scales that do not depend on Lundquist number, it is natural to
imagine that the ``recursive'' reconnection discussed in section~\ref{discussion} should impact the energy cascade in evolving MHD turbulence. A number of recent papers have developed phenomenologies addressing this aspect  \citep{loureiro_2017,  boldyrev_2017, mallet}. Although these works ultimately lead to different predictions for the turbulent cascade, all of them start from an anisotropic MHD turbulence framework and essentially look for the scale at which the elongated eddies, that they associate with current sheets, become unstable to tearing with growth rates equal to those of the nonlinear cascade rate at the same scale. The energy cascade is then modified and sub-inertial spectral power laws has been derived. Here we consider instead how the onset of recursive tearing within current sheets that form {\it between} the eddies may modify the turbulent cascade.

It has been observed in some MHD and kinetic simulations that a power law for the spectral energy density forms when magnetic islands are seen to grow in correspondence of thin current sheets  (e.g., \cite{dong, cerri, franci}). Such findings  support the general idea that the onset of reconnection in an evolving turbulent cascade  has an impact on the cross-scale energy transfer. In the context of kinetic simulations, \cite{slandi} point out that traditional wave  and intermittency based theories fail to explain the observed (steeper) spectra at sub-ion scales, and suggest that the onset of reconnection might be the cause.

The problem of how the evolution of intermittent structures influences the energy cascade is a complex one and far from been understood completely. For example, current sheets might be thought of either as elongated turbulent eddies or as intermittent, coherent structures with depleted nonlinearities forming at boundaries between such eddies. In that case, the phenomenology would require an explicit consideration of eddy filling factors. Here we suggest a possible phenomenological model that describes the feedback of reconnecting intermittent structures on the MHD turbulent cascade, inspired by the intuitive $\beta$-model put forward by \cite{frisch}. 

The idea underlying the $\beta$-model is that while the time scale for nonlinear interactions remains the eddy turnover time (e.g., $\tau_{nl,\ell}\sim  \ell/B_{\ell} $), small eddies become more and more sparse (``less filling''~\cite{frisch}) and will occupy smaller and smaller volumes. A way to include  this effect statistically is to assume that such structures occupy a fraction $\beta_{\ell}$ (the filling factor) of the overall volume, so that the energy density at scale $\ell$ will be given by $E_{\ell}\sim (B_\ell)^2\beta_\ell$,  where $\beta_\ell\leq1$. Since the average energy density transfer rate $\varepsilon$ has to be maintained steady and constant, this typically leads to a steeper spectrum than what predicted traditionally. The usual procedure is to find the scaling of $(B_\ell)^2$ via the condition $\varepsilon\sim E_\ell/\tau_{nl, \ell}\sim (B_\ell)^2\beta_{\ell}/\tau_{nl, \ell}\rightarrow (B_\ell)^2\sim \ell^{2/3}(\beta_\ell)^{-2/3}$, and then go to the continuum to get the spectral density, the whole point being to estimate the filling factor at scale $\ell$.

Consider now a turbulent state that generates elongated 2D current sheets in a plane perpendicular to the local mean magnetic field. Assume that turbulence is magnetically dominated, i.e. neglect the effect of flows on the tearing onset (in reality current sheets are often associated with vorticity sheets so that the flows structure may play a role). Beyond a certain scale $\ell_*$ (that  we leave undetermined at this level of discussion) the  turbulent cascade proceeds inside the localized sheets. In this scenario, the filling volume correction applies when ``recursive'' reconnection develops inside such sheets, and we can estimate $\beta_\ell$ with the volume of magnetic islands times the current sheet number~$\mathcal N$. Thus $\beta_\ell\sim(\ell \delta_\ell) \mathcal{N}$, where  $\ell$ and $\delta_\ell$ are the magnetic island length scales along and across the local 2D sheet, and they correspond to the inverse of our $k$ and $k_\bot$. Now, the scalings given in their general form in eq.~(\ref{scaling_p}) are defined with $L$ being the length scale of the unstable sheet (or  $\ell_*$ with the present notation), which in this case is not necessarily  the macroscopic length, that we label instead $ L_0$. We therefore need to renormalize scales and Lundquist number to their macroscopic values,  $L_0$ and $ S_0$, respectively. This simple procedure leads~to 
\begin{equation}
\beta_\ell\sim  \mathcal{N}  S_0^{-\frac{(1+p)}{(2 + 3 p)}}  \left(\frac{\ell}{ L_0} \right)^{\frac{(3 + 5 p)}{(2+3p)}}.
\label{beta}
\end{equation}

The number of current sheets can be estimated by considering that the dissipation rate must not depend on $S_0$, i.e., $\mathcal N  \eta J^2\sim  (S_0)^0$. Since we expect that current sheets going unstable have an aspect ratio that scales with $S_0$ as $\ell_*/a\sim (S_0)^{1/3}$ or, for generic profiles, $\ell_*/a\sim(S_0)^{1/2(1+p)/(1+2p)}$, we obtain $J^2\sim (S_0)^{(1+p)/(1+2p)}$ and as a consequence $\mathcal N \sim(S_0)^{p/(1+2p)}$ .  If we start from a Kolmogorov-type spectrum, this model predicts a modified energy cascade with a steeper spectral index,  
\begin{equation}
E(k)\sim  \mathcal C S_0^{a}(k L_0)^{-\frac{(13 + 20 p)}{(6 + 9 p)}},
\label{beta2}
\end{equation}

where $\mathcal C$ is a constant that can be determined by matching the cascade in the inertial range with the sub-inertial range, and $-0.02\leq a<0.05$  for $1\leq p< \infty$, and $a=0$ for $p\simeq 1.6$.  

To summarize, eq.~(\ref{beta2}) yields an energy spectrum which is bounded between  $E(k)\sim k^{-11/5}$, for $p=1$, and $E(k)\sim k^{-20/9}$, for $p\rightarrow\infty$. The present model therefore predicts a magnetic energy spectral index of essentially ${-2.2}$.  Inclusion of the Alfv\'en effect as in the Iroshnikov-Kraichnan model~\citep{iroshnikov,kraichnan} is straightforward and would lead to a similar result, with a spectral power law of -2.3.  These results are in good agreement with numerical simulations at high Reynolds and Lundquist number by \cite{dong}, who performed simulations of 2D MHD turbulence; it is also in good agreement with 3D numerical results by~\cite{huang_2016} of an initially unstable Sweet-Parker sheet that develops into a fully turbulent state with spectral index between $-2.3$ and~$-2.1$. Note however that early 2D MHD  simulations of incompressible turbulence by~\cite{biskamp_1989},  and also the more recent large simulations discussed in~\cite{wan_2013},  show the generation of magnetic islands and X-points inside coherent sheets, but in neither case is there evidence of a sub-inertial range. Rather, \cite{wan_2013}, although with a smaller resolution than~\cite{dong}, display a significantly longer inertial range in $k$-space. It would be of interest to investigate further what determines such a different evolution. 

\section{Summary and conclusions}
We have studied via 2.5D MHD simulations the evolution of a thin current sheet unstable to the ``ideal'' tearing mode, by focusing on its nonlinear dynamics. We have shown that the recursive generation of X-points, current sheets and magnetic islands during the nonlinear stage allows to convey energy across scales towards the small, dissipative ones, giving rise to a  turbulent state that can be described by well defined power-law spectra: at the neutral line, magnetic energy displays a power-law spectrum scaling as $B(k)^2\sim k^{-4/5}$ while kinetic energy is much steeper and its spectrum scales as $U(k)^2\sim k^{-2}$, where $k$ is the wave vector along the sheet. Far from the neutral line, a standard Kolmogorov spectrum is approached, both in magnetic and kinetic energy, which  show a scaling very close to $ k^{-5/3}$.   

We have adopted a heuristic method based on the ``recursive'' reconnection already discussed in~\cite{anna, anna2} to provide a model that predicts magnetic energy density spectral power laws which are consistent with our numerical results. We have extended such model to generic current sheet profiles by introducing the parameter~$p$, where hypothetically $1\leq p<\infty$; $p=1$ is the most natural choice as it corresponds to the Harris profile. Our model predicts a  magnetic energy spectrum (for the reconnecting component) which is bound between $B(k)^2\sim k^{-4/5}$ and $B(k)^2\sim k^{-2/3}$, and which undergoes an anisotropic cascade in $k$-space described by $k_\bot\sim k^{3/5}$ and $k_\bot\sim k^{2/3}$, respectively, where $k_\bot$ is the wave vector across the sheet. Remarkably, the spectral power-law index of magnetic energy density and anisotropy are independent of the current sheet thickness and   depend only weakly on the current sheet profile.

We have finally proposed a phenomenological model to incorporate the effect of ``recursive'' reconnection within the turbulent energy cascade, by assuming that fast reconnecting sheets contribute in the form of intermittent structures, as suggested by~\cite{slandi} in a different context. By adapting the $\beta$-model for intermittency~\citep{frisch} to our recursive cascade of magnetic islands, we find that  reconnecting sheets would lead to a modified inertial range with power-law index $-2.2$ when starting from a Kolmogorov state.  

Within the context of turbulence, the present approach neglects the effect of flows (vorticity sheets) on the disruption of turbulent eddies due to  the tearing instability. Some recent discussion on the effect of flows on tearing mode can be found for example in \cite{loureiro_2013}, \cite{anna} and \cite{shi_2018}. The instability of combined vortex-current sheets is a vast topic~\citep{dahlburg}. A generalized model that includes such effects may be considered in future work. It is worth to mention that this work is confined to the two-dimensional geometry, and it will be important to investigate how our results change in three dimensions. Recent simulations have shown that plasmoid chains do not form in three dimensional simulations, although a pressure balance current sheet is used rather than a force-field configuration~\citep{takamoto}. We argue  that the three-dimensional evolution will depend in general on the guide field strength. Three dimensional turbulence simulations within a reduced-MHD model  have shown that elongated quasi two-dimensional current sheets are naturally formed~\citep{rappazzo_2013}, providing support to our two-dimensional model. 

\section*{Acknowledgements}
This research was supported by the NSF-DOE Partnership in Basic Plasma Science and Engineering award n. 1619611 and the NASA Parker Solar Probe Observatory Scientist grant NNX15AF34G. This work used the Extreme Science and Engineering Discovery Environment (XSEDE) Comet at the San Diego Supercomputer Center through allocation TG-AST160007. XSEDE is supported by National Science Foundation grant number ACI-1548562. We also acknowledge the Texas Advanced Computing Center (TACC) at The University of Texas at Austin for providing HPC resources that have contributed to the research results reported within this paper. URL: http://www.tacc.utexas.edu.

\end{document}